\documentclass[prxquantum,twocolumn,preprintnumbers,superscriptaddress,nofootinbib]{revtex4-2} 
\usepackage{graphicx}
\usepackage{bm}
\usepackage{amsmath}
\usepackage{amssymb}
\usepackage{amsfonts}
\usepackage{latexsym,epsfig,bbm}
\usepackage{algorithm}
\usepackage{algorithmic}

\newcommand{\bra}[1]{\langle{#1}|}
\newcommand{\ket}[1]{|{#1}\rangle}

\newcommand{\braket}[2]{\langle{#1}|{#2}\rangle}
\newcommand{\bopk}[3]{\langle{#1}|{#2}|{#3}\rangle}

\newcommand{\del}{\nabla}

\newcommand{\figref}[1]{Fig.~\ref{#1}}

\usepackage{color}
\definecolor{blue}{rgb}{0,0.2,1}

\definecolor{red}{rgb}{0.9,0,0}

\newcommand{\past}[1]{\overleftarrow{#1}}
\newcommand{\fut}[1]{\overrightarrow{#1}}

\newcommand{\tpasto}{t_{\overleftarrow{0}}}
\newcommand{\Tpasto}{T_{\overleftarrow{0}}}
\newcommand{\tfuto}{t_{\overrightarrow{0}}}

\begin{document}

\title{Quantum coarse-graining for extreme dimension reduction\\ in modelling stochastic temporal dynamics}

\author{Thomas J.~Elliott}
\email{physics@tjelliott.net}
\affiliation{Department of Mathematics, Imperial College London, London SW7 2AZ, United Kingdom}

\date{\today}

\begin{abstract}
Stochastic modelling of complex systems plays an essential, yet often computationally intensive role across the quantitative sciences. Recent advances in quantum information processing have elucidated the potential for quantum simulators to exhibit memory advantages for such tasks. Heretofore, the focus has been on lossless memory compression, wherein the advantage is typically in terms of lessening the amount of information tracked by the model, while -- arguably more practical -- reductions in memory dimension are not always possible. Here we address the case of lossy compression for quantum stochastic modelling of continuous-time processes, introducing a method for coarse-graining in quantum state space that drastically reduces the requisite memory dimension for modelling temporal dynamics whilst retaining near-exact statistics. In contrast to classical coarse-graining, this compression is not based on sacrificing temporal resolution, and brings memory-efficient, high-fidelity stochastic modelling within reach of present quantum technologies.
\end{abstract}
\maketitle 

\section{Introduction}

Everywhere we look, we are surrounded by complex systems. They manifest across all scales, from the microscopic level of chemical and physical interactions, through biological processes, to geophysical and meteorological phenomena and beyond~\cite{zurek1990complexity, badii1999complexity, arthur1999complexity, gell2002complexity, ottino2004engineering, grimm2005pattern, boccara2010modeling, crutchfield2012between}. As the descriptor \emph{complex} suggests, with such systems manifesting a rich tapestry of emergent behaviours it quickly becomes an insurmountable task to track their many interacting components in full. Computational tractability demands that when modelling complex systems we keep only a partial knowledge, sufficient for predicting relevant properties of interest. Meanwhile, the remaining information that is discarded (or was not possible to observe in the first place) manifests as stochastic effects on top of this. Accordingly, stochastic modelling~\cite{levinson1986continuously, rabiner1989tutorial, kulp1996generalized, tino1999extracting, palmer2000complexity, yu2002hidden, gerstner2002spiking, bonabeau2002agent, clarke2003application,  bulla2006stylized, park2007complexity, li2008multiscale, wilkinson2009stochastic, haslinger2010computational, yu2010hidden, smouse2010stochastic, garavaglia2011earthquake, kelly2012new} is a critical part of modern science, and identifying ways and means of maximising its efficacy is a transdisciplinary endeavour.

A key bottleneck is the amount of memory available, restricting the amount of information that can be stored. Each configuration the system can take is assigned to a state in the memory; the number of states the memory can support -- its \emph{dimension} -- limits the number of distinct configurations that can be tracked. A form of compression to mitigate this is coarse-graining -- grouping together configurations that are sufficiently close into a single combined configuration, reducing the effective dimension, at the cost of precision. This is particularly prominent for temporal information: time is a continuous parameter requiring an unbounded amount of information to specify to arbitrary precision~\cite{marzen2017informational}; in practice it is coarse-grained into bins of finite width~\cite{marzen2015informational}. 

For a \emph{quantum} memory, the dimension is no longer synonymous with the number of different possible states it can support. In the context of stochastic modelling, by encoding configurations with partially overlapping features into linearly-dependent quantum states, a dimensional compression can be achieved~\cite{thompson2018causal, liu2019optimal, loomis2019strong, elliott2020extreme, ghafari2019dimensional}. This quantum compression advantage can be of significant magnitude~\cite{elliott2020extreme}, though present techniques are constrained to exact (lossless) compression, hampering widespread applicability. Nevertheless, quantum encodings have been shown to almost universally reduce the information cost of stochastic modelling~\cite{gu2012quantum, mahoney2016occam, aghamohammadi2018extreme, elliott2018superior, binder2018practical, elliott2019memory, liu2019optimal, elliott2021memory}, suggesting that many of the dimensions in the memory are barely utilised. This substantiates a strong motivation to develop lossy quantum encodings that trim down these excess dimensions whilst retaining high fidelity with the exact model.

Here we introduce such a lossy compression protocol that can be applied to greatly reduce the memory dimensions devoted to tracking temporal information. Our compression is based on reconstructing approximate -- yet near-exact -- models of a process where the quantum memory states are constrained to a low-dimensional Hilbert space, emancipating the dimension from the number and width of time bins. After reviewing the necessary background, we describe our protocol in detail for pure temporal dynamics, with examples to illustrate the high-fidelities and extreme quantum advantages that can be achieved with only a few memory qubits. We then describe how the protocol can be used for compressed modelling of general continuous-time stochastic processes.

\section{Framework}

\subsection{Stochastic processes and models}

Herein we are concerned with continuous-time, discrete-event stochastic process~\cite{marzen2017structure, elliott2019memory}. These consist of a series of events described by a sequence of couples ${\bm x}_n:=(x_n,t_n)$, where $x_n\in\mathcal{X}$ is the $n$th event in the series and $t_n\in\mathbb{R}^+$ is the time between the $(n-1)$th and $n$th events~\cite{khintchine1934korrelationstheorie}. The sequence is probabilistic, drawn from a distribution \mbox{$P(\ldots,\bm{X}_{n-1},\bm{X}_n,\bm{X}_{n+1},\ldots)$};  throughout we use upper case to represent random variables, and lower case the corresponding variates. We assume the set of possible events $\mathcal{X}$ to be finite. A continguous block of the sequence is denoted \mbox{$\bm{x}_{j:k}:=\bm{x}_j\bm{x}_{j+1}\ldots \bm{x}_{k-1}$}. We consider bi-infinite length sequences such that $n\in\mathbb{Z}$, and assume the process to be stationary such that $P(\bm{X}_{0:L})=P(\bm{X}_{m:m+L})\forall m,L\in\mathbb{Z}$. We will also consider discrete-time approximations to such processes, where times are coarse-grained into finite intervals of size $\Delta t$, recovering the continuous case in the limit $\Delta t\to0$.

We can partition the process into a past and future, delineating what has happened and what is yet to happen respectively relative to some point in the sequence. Without loss of generality we can set $n=0$ to represent the present with $x_0$ the next event to occur, such that the past consists of \mbox{$\past{\bm{x}}:=\bm{x}_{-\infty:0}(\emptyset,\tpasto)$} and the future \mbox{$\fut{\bm{x}}:=(x_0,\tfuto)\bm{x}_{1:\infty}$}. Here, $\tpasto$ represents the time since the last event and $\tfuto$ the time until the next event ($t_0=\tpasto+\tfuto$), and $\emptyset$ denotes that the $0$th event is yet to occur~\cite{marzen2017structure, elliott2019memory}.

We desire models that are able to track relevant information from the past of a process in order to faithfully replicate the corresponding future statistics~\cite{crutchfield1994calculi, crutchfield2012between}. We require the models to be causal, entailing that they can be initialised for any given past, and store no information about the future that could not be obtained from the past observations~\cite{thompson2018causal}. Such models function by means of an encoding function $f:\past{\bm{\mathcal{X}}}\to\mathcal{M}$ that maps pasts into memory states $\rho_m\in\mathcal{M}$, and a transition structure $\Lambda:\mathcal{M}\to\mathcal{M},\emptyset\cup\mathcal{X}$ that produces the future statistics and updates the memory state accordingly~\cite{shalizi2001computational}. In the continuous-time setting this transition structure is a continuous evolution, while in the discrete-time setting it acts once at each timestep~\cite{marzen2015informational, marzen2017informational, marzen2017structure, elliott2018superior}. A model with a lossless encoding is able to replicate the future statistics perfectly, while a lossy one produces an approximation thereof.

\begin{figure}
\includegraphics[width=0.9\linewidth]{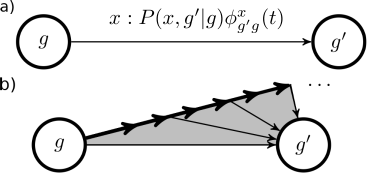}
\caption{(a) HSMM representation of a continuous-time, discrete-event stochastic processes showing the transition structure between modes. Each node corresponds to a mode of the model, and the arrows labelled $x:p(t)$ denote transitions between modes accompanied by event $x$ occuring at time $t$ since the previous event, with the transition occuring with probability $p(t)$. \mbox{(b) Unpacking} into HMM tracking mode occupation times; the nodes continue to represent modes and thin lines the transitions, while the thick black line indicates a continuum of states of the model, tracking both the current mode and time since last event.}
\label{fighmm}
\end{figure}

Continuous-time stochastic processes can be represented by edge-emitting hidden semi-Markov models (HSMMs)~\cite{yu2010hidden, marzen2017structure}. A HSMM comprises of (hidden) modes $\mathcal{G}$, event alphabet $\mathcal{X}$, and transition dynamic $\Lambda$. Conditional on the current mode and the time it has been occupied, the transition dynamic describes the probability of the model emitting a symbol $x\in\mathcal{X}$ and transitioning to a new mode, with the probabilities depending on the particular process [\figref{fighmm}(a)]. 
That is, the system resides in a given mode $g\in\mathcal{G}$ until an emission $x\in\mathcal{X}$ occurs, at which point it transitions to a new mode $g'\in\mathcal{G}$; the probability that a system resides in mode $g$ for a time $t$ before emitting symbol $x$ and transitioning to mode $g'$ is given by the modal wait-time distribution $\sum_{xg'}P(x,g'|g)\phi_{g'g}^x(t)$, where the probabilities $P(X,G'|G)$ describe the symbolic transition structure between modes, and the dwell functions $\phi_{g'g}^x(t)$ the distribution for the time spent in a given mode before such a given transition occurs. See Refs.~\cite{marzen2017structure, elliott2019memory} and Section \ref{secgeneral} for further details.

A HSMM can be unpacked~\cite{elliott2019memory} into an edge-emitting hidden Markov model (HMM)~\cite{rabiner1986introduction} with a continuous-state space tracking the occupation time for the modes [\figref{fighmm}(b)]. States in the HMM represent a mode and time since last event $(g,\tpasto)$, with a transition structure taking the system to $(g,\tpasto+dt)$ on non-events in the next infinitesimal time interval $dt$, and $(g',0)$ upon events. The corresponding emitted symbols are $\emptyset$ for non-events and $x\in\mathcal{X}$ for each event; transition probabilities follow from the conditional form of the modal wait-time distributions. Models of discrete-time stochastic processes can similarly be modelled by discrete-state HMMs, in which the occupation time is tracked by the corresponding coarse-grained states~\cite{marzen2015informational}.

\subsection{Memory and quantum advantage}

A key metric of efficiency for a model is how much memory it requires to operate~\cite{garner2017provably}. One way this can be parameterised is the information cost -- in the sense of Shannon entropy -- of storing the compressed past information~\cite{shalizi2001computational, gu2012quantum, garner2017provably}. Another, to which we direct our focus here, is the size of the substrate into which this information is encoded -- in other words, the dimension of the memory state space~\cite{shalizi2001computational, yang2018quantum, liu2019optimal, elliott2020extreme, yang2020ultimate}. The choice of encoding function will impact upon the memory cost, and is ideally chosen to make it as small as possible.

For stationary stochastic processes the optimal classical lossless memory encoding function is provided by the causal equivalence relation ($\sim_\varepsilon$) of computational mechanics~\cite{crutchfield1989inferring, shalizi2001computational, crutchfield2012between}, which partitions the entire set of semi-infinite pasts $\past{\bm{\mathcal{X}}}$ into equivalence classes called causal states $s\in\mathcal{S}$ such that two pasts belong to the same causal state iff they effect the same conditional future statistics:
\begin{equation}
\label{eqcausalequiv}
P(\fut{\bm{X}}|\past{\bm{x}})=P(\fut{\bm{X}}|\past{\bm{x}}')\Leftrightarrow \past{\bm{x}}\sim_\varepsilon\past{\bm{x}}'.
\end{equation}
The memory-optimal lossless classical model (known as the $\varepsilon$-machine) is then constructed by designating a memory state $\ket{s}$ for each causal state $s$, and having the causal state encoding function $f_\varepsilon$ assign pasts accordingly. A typical process evolving in continuous-time will require an infinite-dimensional memory to record the progress through infinitesimal divisions in time~\cite{marzen2017informational, elliott2018superior, elliott2020extreme}, engendering the need for lossy approximations that evolve with discretised timesteps~\cite{marzen2015informational, elliott2020extreme}.

With the advent of quantum information processing tools, the optimality of $\varepsilon$-machines has been supplanted~\cite{gu2012quantum}. Quantum encoding functions $f_q$ map pasts into a set of quantum memory states; by leveraging the possibility of encoding information into an ensemble of non-orthogonal states, further compression beyond the causal state encoding function may be attained. Prior work has centred on lowering the information cost of storing the past~\cite{gu2012quantum, mahoney2016occam, aghamohammadi2018extreme, elliott2018superior, binder2018practical, elliott2019memory, liu2019optimal, elliott2021memory}, showing that a quantum compression advantage can almost always be procured. Recent focus has been devoted to obtaining corresponding advantages in compressing the dimension of the memory, by engineering quantum memory states with linear dependencies~\cite{thompson2018causal, liu2019optimal, loomis2019strong, elliott2020extreme, ghafari2019dimensional}. Examples have highlighted that such dimensional compression can sometimes be made arbitrarily strong with respect to the optimal classical encoding~\cite{elliott2020extreme}, though instances where it may be achieved in the lossless regime appear to be much less ubiquitous than in the case of reducing the information cost~\cite{liu2019optimal}. The lossy encoding protocol we introduce seeks to remedy this present shortcoming of the quantum models in the context of tracking the temporal aspect of their dynamics, to escape the associated memory dimension divergence in the continuous limit.

\subsection{Renewal processes}

With our attention directed towards compressing the temporal information, for much of this manuscript we will work with a special class of continuous-time stochastic processes that are purely temporal in nature: renewal processes~\cite{smith1958renewal}. These consist of a single mode and a single symbol, such that the resulting process is a series of identical events stochastically separated in time, with the spacing of each consecutive pair of events drawn from the same distribution. The distribution governing the time between events is called the wait-time distribution $\phi(t)$, and the survival probability $\Phi(t):=\int_t^\infty\phi(t')dt'$ is the probability that a given interval is of length $t$ or greater~\cite{marzen2015informational, marzen2017informational, elliott2018superior, elliott2020extreme}.

With few exceptions, for generic renewal processes the causal states group pasts together according to the time since the last event occurred~\cite{marzen2015informational, marzen2017informational, elliott2018superior}. That is, all relevant information for predicting the future of a renewal process is contained within the time since last event -- such that the causal states are in one-to-one correspondence with $\tpasto$ -- and moreover, can only provide predictive power with respect to the time $\tfuto$ until the next event will happen. 

\begin{figure}
\includegraphics[width=\linewidth]{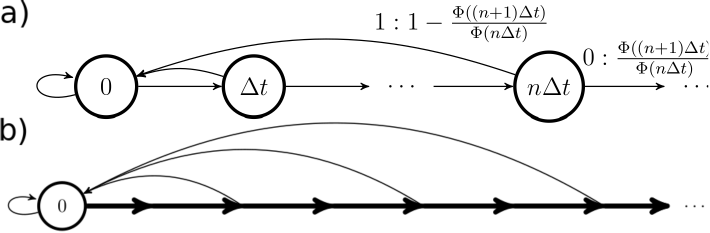}
\caption{(a) Discrete and (b) continuous HMM representations of $\varepsilon$-machines of a renewal process. The system progresses along a counter until an event occurs, upon which it transitions to the reset state. In (a) nodes correspond to states of the HMM tracking time since last event, in (b) the thick black line represents a continuum of such states. The thin arrows represent transitions between states, with $x:p$ indicating the probability $p$ of the transition occuring, accompanied by symbol $x$. Symbol 1 represents events, and 0 non-events.}
\label{figrenewalhmm}
\end{figure}

The transition structure between the memory states of the $\varepsilon$-machine for a renewal process has been likened to a `conveyor belt'~\cite{marzen2017informational}, progressing continuously along a line with time until an event occurs, whereupon the memory jumps to a `reset' state corresponding to $\tpasto=0$. The probability of occupying the memory state corresponding to $\tpasto$ is given by $\pi(\tpasto)=\mu\Phi(\tpasto)$, where \mbox{$\mu:=\left(\int_0^\infty t\phi(t)dt\right)^{-1}$} is the so-called mean firing rate~\cite{marzen2017informational, elliott2018superior}. The discrete-time analogue consists of a linear sequence of memory states through which the system progresses, akin to the incrementation of a counter, until also resetting upon an event~\cite{marzen2015informational, elliott2020extreme}. Both are illustrated in \figref{figrenewalhmm}. The exact continuous-time version requires an infinite continuum of memory states, and thus requires a memory of unbounded dimension; when there is no maximum value for $\tpasto$ the discrete-time case will similarly need an infinite-dimensional memory, and thus finite-dimensional approximations must also adopt a terminal state that the counter cannot exceed~\cite{elliott2020extreme}.

\section{Quantum coarse-graining}

\subsection{Quantum models of renewal processes}

In previous work~\cite{elliott2018superior} we have established that a general renewal process with wait-time distribution $\phi(t)$ can be exactly simulated by a quantum model with a memory encoding function $f_q(\past{\bm{x}})=\ket{\varsigma_{\tpasto}}$, where
\begin{equation}
\label{eqquantumrenewalstates}
\ket{\varsigma_{\tpasto}}:=\int_0^\infty\frac{\psi(\tpasto+t)}{\sqrt{\Phi(\tpasto)}}dt\ket{t},
\end{equation}
with $\{\ket{t}\}$ an infinite-dimensional orthogonal basis and $\psi(t):=\sqrt{\phi(t)}$.\footnote{Note that in principle an arbitrary, time-dependent complex phase can be added to $\psi(t)$; provided that $|\psi(t)|^2=\phi(t)$ the model will still yield the correct statistics, albeit with a potentially different memory cost. We return to this point later.} The future statistics are extracted from these memory states by means of a continuous measurement sweep that at each infinitesimal interval $\delta t$ produces a binary outcome as to whether or not the system is found in a state $\ket{t}$ in the interval $[0,\delta t)$: if yes, then the event is deemed to have occured and the memory is re-initialised in state $\ket{\varsigma_0}$; if not then the event does not occur, and a relabelling $t\to t-\delta t$ takes place.

A fine-grained discrete analogue of this evolution with time-step interval $\delta t$ can be implemented through the following unitary interaction $U_{\delta t}$ coupling the memory state to an ancillary system used to provide the measurement readout, where 0 and 1 represent non-events and events respectively~\cite{elliott2020extreme}:
\begin{equation}
\label{eqrenewalu}
U_{\delta t}\ket{\varsigma_t}\ket{0}\!=\!\sqrt{\frac{\Phi(t\!+\!\delta t)}{\Phi(t)}}\ket{\varsigma_{t+\delta t}}\ket{0}\!+\!\sqrt{1\!-\!\frac{\Phi(t\!+\!\delta t)}{\Phi(t)}}\ket{\varsigma_0}\ket{1}.
\end{equation}
After measurement, the ancilla is set to $\ket{0}$ ready for the next timestep. The amplitudes on the right-hand side of this equation are set such that they yield the correct probabilities for the future statistics, as \mbox{$\int_t^{t+\delta t}\phi(t')dt'=\Phi(t)-\Phi(t+\delta t)$}. Arbitrary complex phases can be added to these amplitudes without affecting the statistics~\cite{liu2019optimal, elliott2020extreme}; on the first term it is equivalent to appending an irrelevant phase to the quantum memory states, while on the latter it mirrors the effect of a complex phase on $\psi(t)$.

\subsection{Quantum model memory as an integral kernel}

The steady-state $\rho$ of the quantum model memory is given by a mixture of the quantum memory states, weighted by their probability of occurence~\cite{elliott2018superior}:
\begin{align}
\label{eqquantummem}
\rho:=&\int_0^\infty \pi(\tpasto)d\tpasto\ket{\varsigma_{\tpasto}}\bra{\varsigma_{\tpasto}}\nonumber\\
=&\mu\iiint_0^\infty\psi(\tpasto+t)\psi(\tpasto+t')dtdt'd\tpasto\ket{t}\bra{t'}.
\end{align}

The rank of $\rho$ corresponds to the dimension required by the memory substrate to support the range of quantum memory states. This is given by the number of non-zero elements in the spectrum of $\rho$, which can be found from the characteristic equation
\begin{equation}
\label{eqchar}
\int_0^\infty\rho(t,t')\nu(t')dt'=\lambda\nu(t).
\end{equation}
This has the form of a homogenous Fredholm integral equation of the second kind~\cite{delves1988computational}, with \mbox{$\rho(t,t')=\mu\int_0^\infty\psi(\tpasto+t)\psi(\tpasto+t')d\tpasto$} corresponding to the kernel of the equation. 

We are thus in a position to leverage results from Fredholm theory to reveal properties of the spectrum $\{\lambda\}$ of $\rho$. Most pertinently, if $\rho$ represents a degenerate kernel, wherein it can be expressed as $\rho(t,t')=\sum_{j=1}^N\alpha_j(t)\beta_j(t')$ for some finite integer $N$ and set of functions $\{\alpha_j,\beta_j\}$, then the spectrum has at most $N$ non-zero elements~\cite{delves1988computational}. Consequently, the memory states can be stored within an $N$-dimensional space. However, the general form of $\rho$ as per Eq.~\eqref{eqquantummem} does not readily present as a degenerate kernel, and indeed, exact quantum models of renewal processes often require an infinite-dimensional memory space. Nevertheless, the amount of information retained in the memory about the past of the process typically appears to be finite~\cite{elliott2018superior}, suggesting many of these dimensions are barely utilised and motivating the pursuit of a lossy -- yet still near-exact -- compression method. A suggestive path to such compression is to truncate $\rho$ by removing the dimensions corresponding to elements of its spectrum that are sufficiently small (as the $\{\lambda\}$ represent the occupation probabilities of the eigenstates of $\rho$). However, this impacts upon the transition structure of the model, rendering it non-physical. An approach with greater finesse is needed, which we now provide.

\subsection{Exponential sums and lossy compression}

Rather than taking an existing exact model and introducing lossy distortion to effect compression, we will instead construct a distortion of the underlying process that is amenable to simulation by a model with a memory of low dimension. The intent is that the exact model of the distorted process forms a near-exact, compressed model of the original process.

This requires us to identify what features the wait-time distribution must possess to permit a finite-dimensional exact model. In other words, to identify what the constraints on $\phi(t)$ are such that it will lead to $\rho(t,t')$ taking the form of a degenerate kernel. Let us begin by introducing the kernel \mbox{$\kappa(t,t'):=\psi(t+t')$}, such that \mbox{$\rho(t,t')=\mu\int_0^\infty\kappa(t,t'')\kappa(t'',t')dt''$}. It then follows that the spectrum of $\kappa(t,t')$ is $\{\sqrt{\lambda/\mu}\}$, and is thus of the same rank as $\rho(t,t')$~\cite{delves1988computational}. This reduces the problem to identifying the conditions under which $\kappa(t,t')$ is a degenerate kernel. These are then the processes for which we can express $\psi(t)$ as a finite sum of functions $F_j(t)$ that satisfy \mbox{$F_j(t+t')=\alpha_j(t)\beta_j(t')$}. We can readily identify the appropriate functions as being (complex) exponentials, i.e., $F_j(t)=c_j\exp(z_jt)$ for some $(c_j,z_j)\in\mathbb{C}^2$. Thus, for \mbox{$\psi(t)=\sum_{j=1}^Nc_j\exp(z_jt)$} we correspondingly have at most $N$ non-zero eigenvalues of the kernel $\kappa(t,t')$. 

Though we began by assuming $\psi(t)$ is real, if we allow it to be complex we instead have \mbox{$\phi(t)=|\psi(t)|^2$}, and \mbox{$\rho(t,t')=\mu\int_0^\infty\psi(t+\tpasto)\psi^*(t'+\tpasto)d\tpasto$}. Notice that even when \mbox{$\psi(t)=\sum_{j=1}^Nc_j\exp(z_jt)$} is complex, it can be verified through direct substitution that $\rho(t,t')$ remains a degenerate kernel of at most rank $N$. Thus, with an $N$-dimensional memory it is possible to model renewal processes for which
\begin{equation}
\label{eqexactdecomp}
\phi(t)=\left|\sum_{j=1}^Nc_je^{z_jt}\right|^2.
\end{equation}
Let us decompose $z_j:=-\gamma_j+i\omega_j$ for $(\gamma_j,\omega_j)\in\mathbb{R}^2$. For $\phi(t)$ to be a valid distribution it must be normalisable to unity, and thus we can constrain $\gamma_j\in\mathbb{R}^+$. 

The complex exponentials $\exp(-zt)$ form an overcomplete basis into which any piecewise continuous function of finite exponential order can be decomposed, where the overlap of the function with the basis elements are described by its Laplace transform. Thus, for any $\psi(t)$ that is piecewise continuous and of finite exponential order we can express the corresponding wait-time distribution in the form of Eq.~\eqref{eqexactdecomp}, albeit with $N$ not necessarily finite. 

Nevertheless, this provides a constructive approach to finding lossy compressions for quantum models of renewal processes. The goal is to find exponential sums with a finite number of terms that provide a high-fidelity approximation to $\psi(t)$. In practice, it has been found that such decompositions can achieve accurate reconstructions of a function with a relatively small number of terms. Moreover, there are systematic approaches to obtaining such decompositions.\footnote{We will not designate any particular such method as the optimal. In testing our protocol we found the algorithm of Beylkin and Monz\'{o}n~\cite{beylkin2005approximation} to perform well with low computational cost.} From the decomposition we are then able to build an exact model of the approximate process, to effect a near-exact model of the original process.

The last step remaining is to find an explicit encoding of the memory states of the approximate model into a finite-dimensional memory space. Beginning from a (normalised) approximate decomposition $\tilde{\psi}(t)=\sum_{j=1}^Nc_j\exp((-\gamma_j+i\omega_j)t)$, we assign $N$ `generator' states $\{\ket{\varphi_j}\}$ and a unitary operator $\tilde{U}_{\delta t}$ with the evolution\footnote{Unlike Eq.~\eqref{eqrenewalu}, there is no freedom in these amplitudes; changing their magnitude is equivalent to changing $\gamma_j$, while phase factors correspond to different $c_j$ and $\omega_j$.}
\begin{equation}
\label{eqlossyu}
\tilde{U}_{\delta t}\ket{\varphi_j}\ket{0}=e^{(-\gamma_j+i\omega_j)\delta t}\ket{\varphi_j}\ket{0}+\sqrt{1-e^{-2\gamma_j\delta t}}\ket{\tilde{\varsigma}_0}\ket{1},
\end{equation}
in analogy with Eq.~\eqref{eqrenewalu}. Here, we have defined 
\begin{equation}
\ket{\tilde{\varsigma}_0}:=\sum_{j=1}^N\frac{c_j}{\sqrt{2\gamma_j}}\ket{\varphi_j},
\end{equation}
which forms the reset state corresponding to $\tpasto=0$, with the rest of the quantum memory states $\{\ket{\tilde{\varsigma}_{\tpasto}}\}$ implicitly defined by acting $U$ with the ancilla a sufficient number of times, postselected on all measurement outcomes being 0, i.e., \mbox{$\ket{\tilde{\varsigma}_{n\delta t}}=\bra{0}(I\otimes\ket{0}\bra{0}U)^n\ket{\tilde{\varsigma}_0}\ket{0}$}. Non-normalised, these states can also be expressed \mbox{$\ket{\tilde{\varsigma}_{\tpasto}}\propto\sum_{j=1}^N(c_j/\sqrt{2\gamma_j})\exp((-\gamma_j+i\omega_j)\tpasto)\ket{\varphi_j}$}. The overlaps of the generator states can be obtained~\cite{binder2018practical, liu2019optimal} from the recursive relations \mbox{$\braket{\varphi_j}{\varphi_k}=\bra{\varphi_j}\bopk{0}{\tilde{U}^\dagger_{\delta t} \tilde{U}_{\delta t}}{\varphi_k}\ket{0}$}, from which we can move from their implicit definition to express them explicitly in terms of an $N$-dimensional set of orthonormal basis states using a reverse Gram-Schmidt procedure~\cite{dennery1996mathematics}. The relevant columns of $\tilde{U}_{\delta t}$ are defined implicitly by Eq.~\eqref{eqlossyu} and can now readily be expressed explicitly in this basis; the remaining columns can be assigned arbitrarily provided they preserve orthonormality of the basis states (by using e.g., a Gram-Schmidt procedure)~\cite{binder2018practical}. 

This constructs a lossy compression of the quantum memory states, yielding a near-exact model of the process. The steps are summarised in Algorithm 1.

\begin{algorithm}[H]
\caption{\textsf{\\\mbox{Quantum coarse-graining for modelling renewal processes}}}
\label{algcoarsegrain}
\begin{flushleft}
\emph{Inputs}: Renewal process wait-time distribution $\phi(t)$. \\
\emph{Outputs}: Compressed quantum memory states $\{\ket{\tilde{\varsigma}_{\tpasto}}\}$, evolution $\tilde{U}_{\delta t}$, approximate wait-time distribution $\tilde{\phi}(t)$. 
\end{flushleft}
\begin{algorithmic}[1]
\STATE Define $\psi(t)=\sqrt{\phi(t)}$.
\STATE Use method of choice to find an exponential sum \mbox{$\tilde{\psi}(t)=\sum_{j=1}^Nc_j\exp((-\gamma_j+i\omega_j)t)$} of $N$ terms approximating $\psi(t)$. Scale weights such that $\tilde{\phi}(t)$ is normalised.
\STATE Implicitly define $\tilde{U}_{\delta t}$ and $\{\ket{\varphi_j}\}$ according to Eq.~\eqref{eqlossyu} and determine the state overlaps from the recursive relations \mbox{$\braket{\varphi_j}{\varphi_k}=\bra{\varphi_j}\bopk{0}{\tilde{U}^\dagger_{\delta t} \tilde{U}_{\delta t}}{\varphi_k}\ket{0}$}. Use a reverse Gram-Schmidt procedure to express the states in terms of $N$ orthonormal basis states.
\STATE Assign columns of $\tilde{U}_{\delta t}$ defined in Eq.~\eqref{eqlossyu}. Fill the remaining columns arbitrarily, using a Gram-Schmidt procedure to ensure orthonormality with existing columns.
\STATE Assign compressed quantum memory states $\{\ket{\tilde{\varsigma}_{\tpasto}}\}$:
\begin{equation*}
\ket{\tilde{\varsigma}_{\tpasto}}\propto\sum_{j=1}^N\frac{c_j}{\sqrt{2\gamma_j}}e^{(-\gamma_j+i\omega_j)\tpasto}\ket{\varphi_j}.
\end{equation*}
\end{algorithmic}
\end{algorithm}

\section{Examples}
\label{secexamp}

As a demonstration of the efficacy of our quantum compression protocol, we apply it to the modelling of two example renewal processes. For each process we show how the quantum models quickly converge on high-fidelity approximations of the original processes with only a comparatively small memory dimension. Our approximate exponential sums are found using the method of Beylkin and Monz\'{o}n~\cite{beylkin2005approximation}, summarised in Appendix \ref{secdecomp}. 

We quantify the goodness-of-fit using a Kolmogorov-Smirnov (KS) statistic~\cite{massey1951kolmogorov}, which is defined as the maximum difference between points in the the cumulative distribution functions of two probability distributions. This allows us to compare how well discrete distributions approximate continuous distributions, as the cumulative distribution function can be extended over a continuum. That is, let $C_p(t)=\int_0^tp(t')dt'$ be the cumulative distribution function of a continuous distribution $p(t)$, and $C_q(t)=\sum_{n=0}^{\mathrm{argmax}(N|N\delta t<t)}q(n\delta t)$ the continuum form of the cumulative distribution of a discrete distribution $q(n\delta t)$. The KS statistic is then given by $\mathrm{KS}(p,q)=\mathrm{max}_t |C_p(t)-C_q(t)|$. For a renewal process the survival probability $\Phi(t)=1-C_\phi(t)$, and so the KS statistic here also corresponds to the maximum difference between the survival probabilities of the exact and approximate processses at any time: $\mathrm{KS}(\phi(t),\tilde{\phi}(t))=\mathrm{max}_t|\Phi(t)-\tilde{\Phi}(t)|$, where $\tilde{\Phi}(t):=\int_t^\infty|\tilde{\psi}(t')|^2dt'$. Thus, the KS statistic as employed here measures the largest cumulative divergence between the statistics of the approximate model and the exact process.

We compare our quantum models to approximate classical models constrained to a classical memory of the same dimension. These classical models are constructed by discretising the process into finite-sized time-steps and using gradient descent~\cite{hastie2009elements} to fit the parameters, taking the KS statistic as a cost function (see Appendix \ref{secclassicalcomp}). While we do not claim this to be the optimal lossy classical compression, we believe it to provide a fair indicator of the potential performance of classical compression methods for this task.

\subsection{Alternating Poisson process}

\begin{figure}
\includegraphics[width=\linewidth]{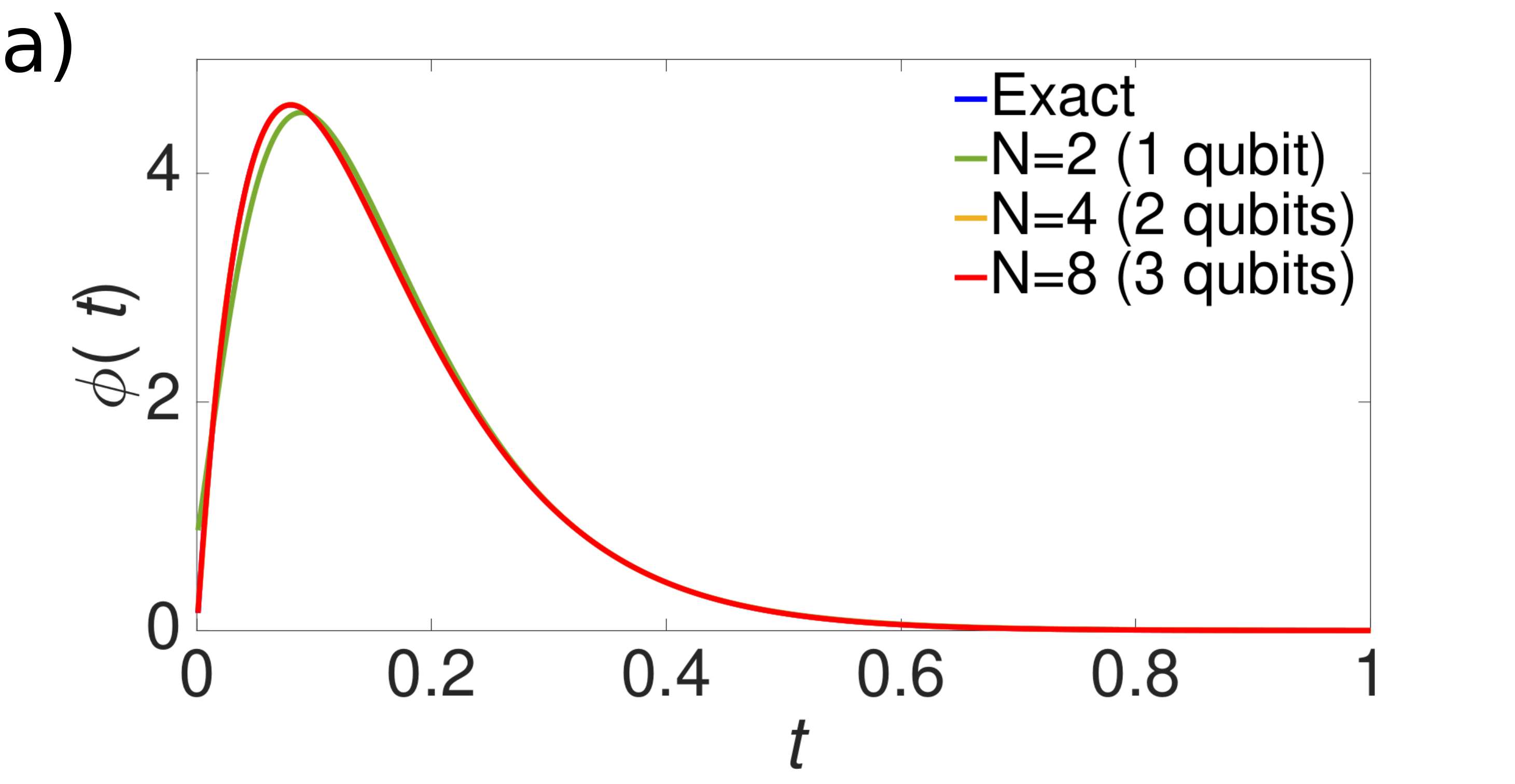}
\includegraphics[width=\linewidth]{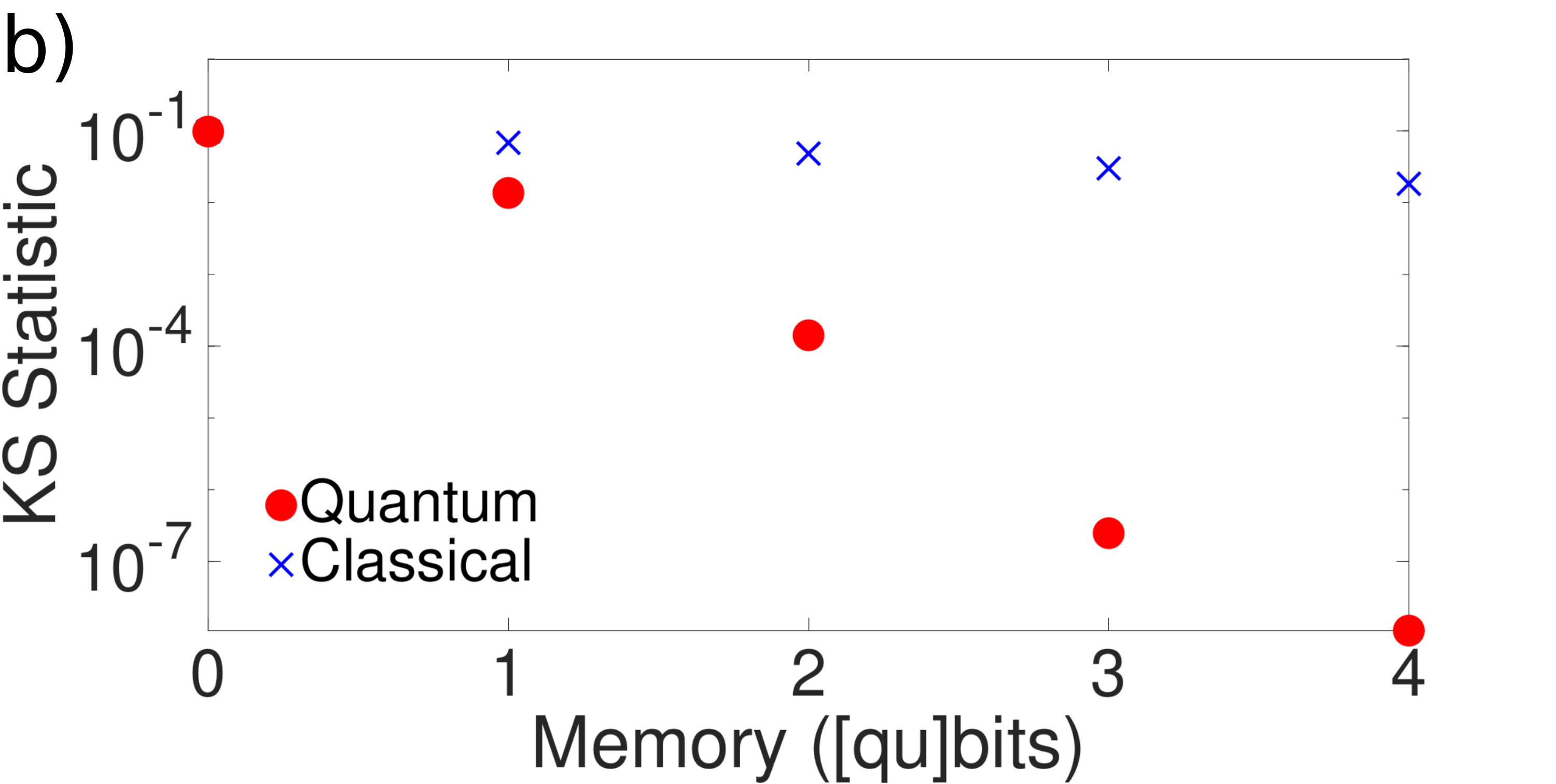}
\caption{(a) Wait-time distributions of compressed quantum models of an alternating Poisson process in arbitrary units. (b) KS statistics comparing performance of compressed quantum models to compressed classical.}
\label{figsns}
\end{figure}

As a first example, we consider an alternating Poisson process. The output can be described by a sequential series of Poisson processes, with an event on these underlying processes alternatively coinciding with events or non-events of the alternating Poisson process (non-events of the Poisson processes also correspond to non-events of the alternating Poisson process). The corresponding wait-time distribution is given by
\begin{equation}
\label{eqappwait}
\phi(t)=\gamma^2te^{-\gamma t},
\end{equation}
where the rate $\gamma$ sets an arbitrary scale for units of time. This is the continuous-time analogue of the so-called simple non-unifilar source process~\cite{crutchfield1994calculi}. While also appearing simple to generate, it too has no finite-dimensional exact causal classical representation~\cite{marzen2015informational}; it is thought that an exact causal quantum model is similarly structurally complex. 

Using our compression protocol, we observe excellent performance in replicating the statistics of the alternating Poisson process with low-dimensional quantum models. As can be seen in \figref{figsns}(a), even a single qubit memory provides a close approximation to the exact wait-time distribution, and a two qubit memory is seemingly indistinguishable at the resolution shown. In \figref{figsns}(b) we compare the perfomance of our coarse-grained quantum models with the classical approximations, as well as a memoryless model. We see that the quantum models bear a KS statistic orders of magnitude smaller than the corresponding classical, and moreover, appear to exhibit a more favourable scaling with increasing memory.

\subsection{Bimodal Gaussian process}

\begin{figure}
\includegraphics[width=\linewidth]{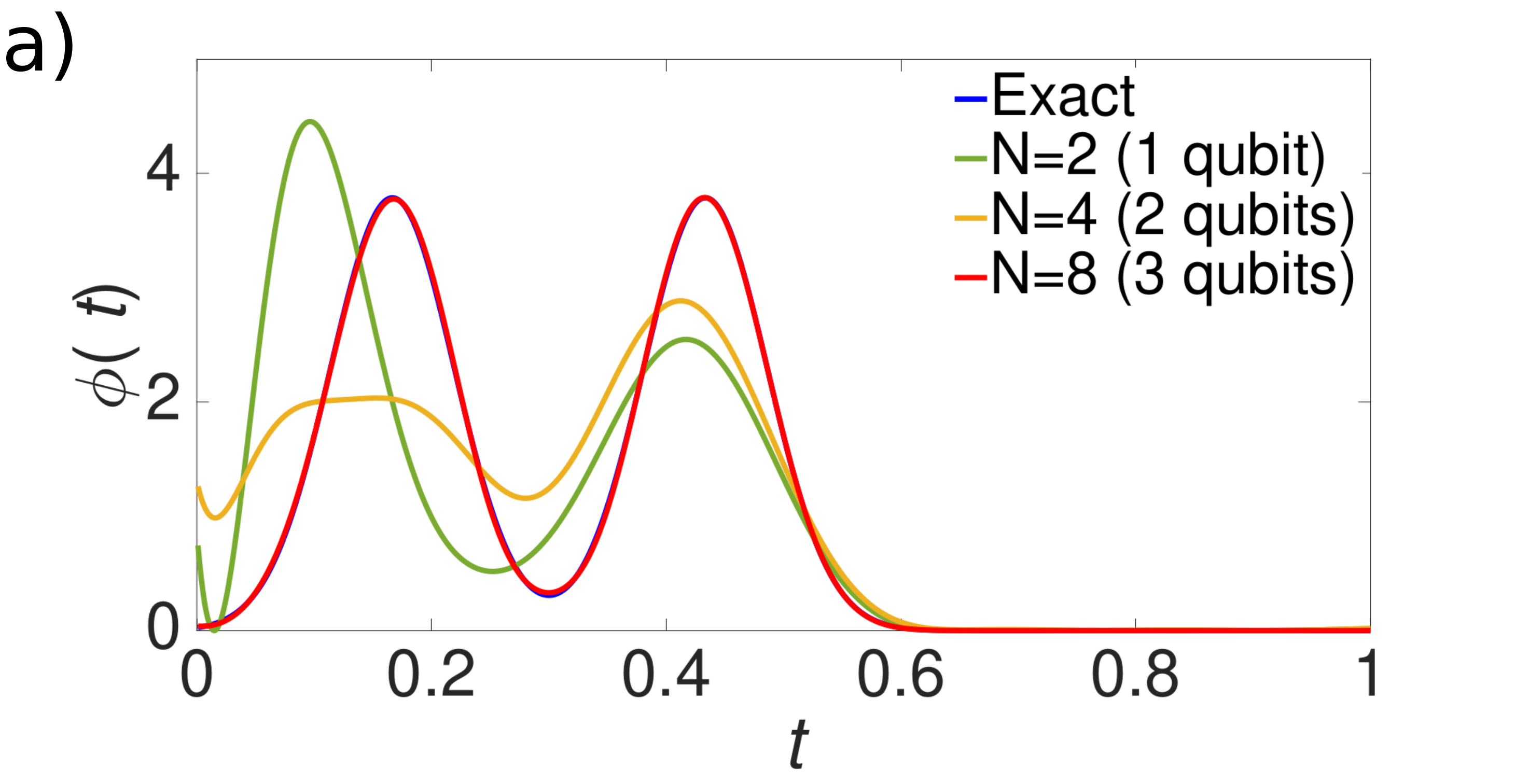}
\includegraphics[width=\linewidth]{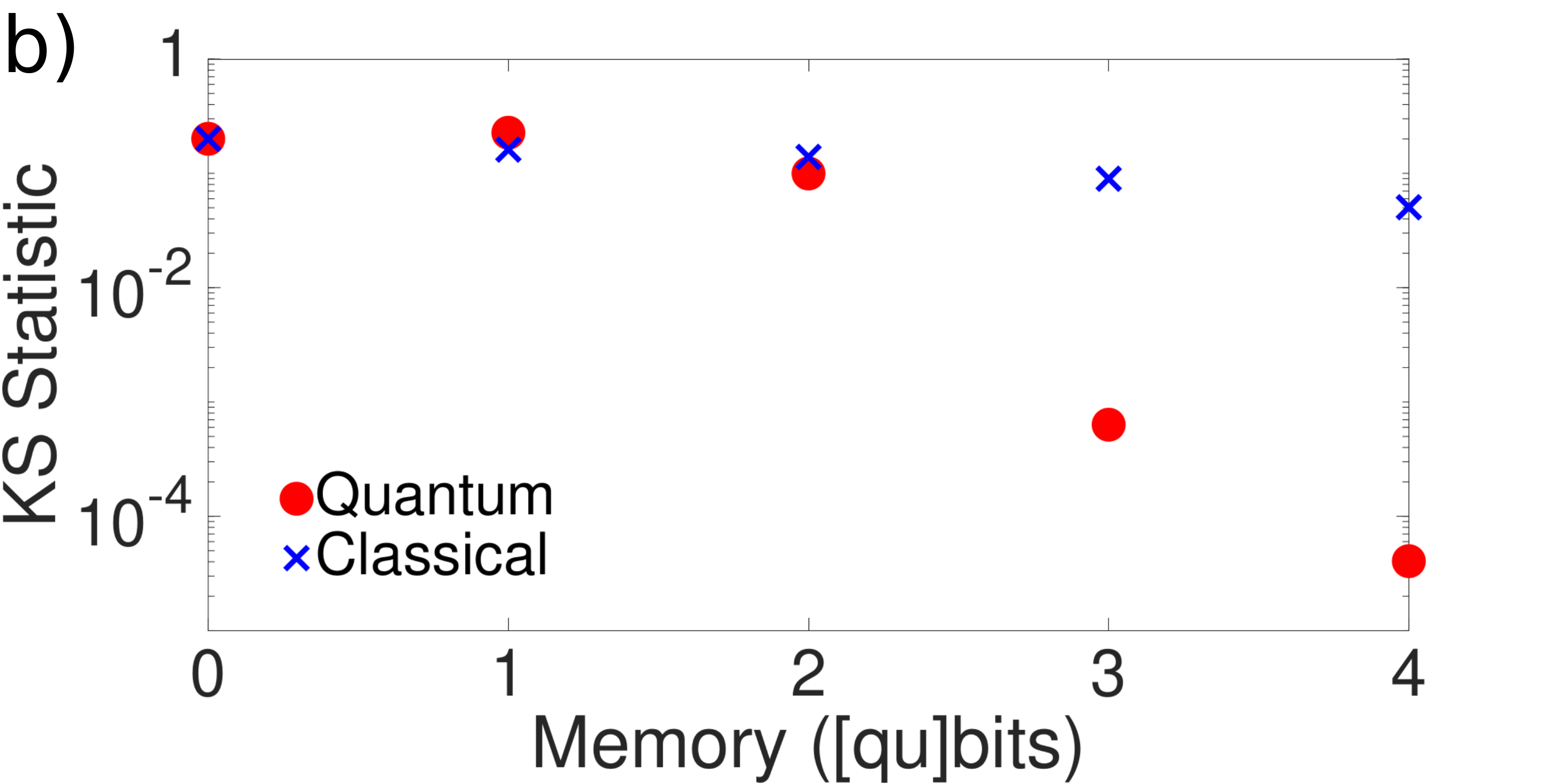}
\caption{(a) Wait-time distributions of compressed quantum models of a bimodal Gaussian process in arbitrary units. (b) KS statistics comparing performance of compressed quantum models to compressed classical.}
\label{figbmg}
\end{figure}

For the second example we find compressed models of a bimodal Gaussian process. The wait-time distribution consists of the sum of two displaced Gaussian peaks:
\begin{equation}
\label{eqbngwait}
\phi(t)=p_1e^{-(t-\mu_1)^2/\sigma_1^2}+p_2e^{-(t-\mu_2)^2/\sigma_2^2}.
\end{equation}
As with the previous example, the units of time are arbitrary, and can be set through the $\sigma$. We consider the case where the two peaks have equal weight ($p_1=p_2$) and equal spread ($\sigma_1=\sigma_2$). In units where $\sigma=1$, we then take $\mu_1=\sqrt{5}$ and $\mu_2=\sqrt{33.8}$. This leads to little overlap between the two peaks, requiring a model to be able to capture features at both short and long timescales in order to account for the two regions of high event probability, and the low probability trough between them.

As can be seen in \figref{figbmg}(a), our coarse-grained models struggle to fully capture the features with one and two qubit memories, with the former overweighting the first peak, and the latter the second. With a three qubit memory however, the model closely follows the exact process. This is reflected in the KS statistic [\figref{figbmg}(b)], where there is a drastic decrease when going from two qubits to three. This is possibly due to the method used here to construct the approximate exponential sum: rather than fixing the maximum allowed number of terms in advance, the method instead constructs a sum with a large number of terms and then afterwards truncates to those with the largest weight. In this case, we find that the terms lost to truncation are not always negligible. This motivates future consideration of alternative methods for constructing approximate exponential sums that begin with the constraint of a maximum allowed number of terms, in order to make best use of the available memory resources. Nevertheless, we still see that our coarse-grained quantum models significantly outperform classical models with only a small number of qubits.

\section{Costly features?}
\label{seccostly}

We have seen that the quantum compression protocol performs well on the two examples above. However, this begs the question of how well it performs in general, and for which processes it will show the weakest perfomance. Ultimately, the accuracy of the model comes down to how good an approximation the finite exponential sum is of the wait-time distribution -- or conversely, the dimension required by the model depends on how few terms are required in the sum to reach a desired precision -- as the compressed model will (experimental imperfections aside) provide an exact model of this approximation of the wait-time distribution. In this sense, the performance of our compression protocol comes down to how well the method used to construct an approximate exponential sum performs. For the particular algorithm used in our examples we refer the reader to the discussion in the associated literature~\cite{beylkin2005approximation, beylkin2010approximation}, also noting that they find even better performance in practice than indicated by their bounds. 

Nevertheless, we can find a useful heuristic in the information cost of the exact quantum model of the process -- once the (logarithm of) the dimension drops below the information cost (i.e., once the capacity of the memory is lower than the information required for exact modelling), the compressed model must throw away useful information, limiting the accuracy it can achieve. Correspondingly, we can expect the performance of the quantum compression to be inversely correlated with the information cost of exact quantum modelling. 

We can also deduce the features that would be most stubborn to compress. Consider our discussion above comparing the exponential sum with expressing the function in the Laplace basis. Given that we want our sum to have as few terms as possible, problematic functions are those that are highly-localised, as they have large spread in the Laplace basis. Indeed, the ultimate limit of this -- $\delta$-functions -- represent deterministic renewal processes; such processes do not allow a quantum advantage even in information cost in exact compression settings~\cite{gu2012quantum, elliott2018superior}. In Appendix \ref{sectophat} we provide a case study of the performance of our quantum compression protocol applied to a series of top-hat wait-time distributions of decreasing width. These processes represent increasingly accurate models of ideal clocks~\cite{woods2018quantum, yang2020ultimate}, and are also similarly difficult for classical compression methods. More generally, processes dominated by such sharp peaks are resistant to quantum compression in the information cost~\cite{elliott2018superior}, and so can be expected to also present difficulties for methods of compressing the memory dimension such as ours.

\section{Deployment with general continuous-time stochastic processes}
\label{secgeneral}

\subsection{Generalising the protocol}

Algorithm 1 -- our protocol for compressing quantum models of renewal processes can be adapted to compress the temporal aspect of quantum models of general continuous-time processes with multiple modes and events~\cite{marzen2017structure, elliott2019memory}. 

Consider such a process with modes $g\in\mathcal{G}$, events $x\in\mathcal{X}$ and a transition dynamic $\Lambda$. The dynamic $\Lambda$ effects an evolution according to $P(X,G'|G,\Tpasto)$ describing the probability density of an event $x$ occuring accompanied by a transition into mode $g'$ in the next infinitesimal interval $dt$ given the system is currently in mode $g$ with time $\tpasto$ since the last event. Following the corresponding literature on memory-minimal classical models~\cite{marzen2017structure} we assume a HSMM representation of the process where the subsequent mode is uniquely determined by $(g,x)$ -- independent of $t_0$. This is a slightly stronger condition than strictly necessary for the model to be causal, and we discuss its relaxation later.

Along with the modal wait-time distributions \mbox{$\sum_{xg'}P(x,g'|g)\phi_{g'g}^x(t)$} we can define a corresponding modal survival probability \mbox{$\Phi_g(t)=\sum_{xg'}\int_t^\infty P(x,g'|g)\phi_{g'g}^x(t')dt'$}~\cite{elliott2019memory}. From these one can then define a set of quantum memory states $\{\ket{\varsigma_{g\tpasto}}\}$ and evolution $U_{\delta t}$ such that\footnote{These exact quantum models generally achieve greater compression in information cost than analogous prior quantum models~\cite{elliott2019memory} as they better compensate for non-Markovianity in the transitions between modes.}
\begin{align}
\label{eqexactgeneralu}
U_{\delta t}\ket{\varsigma_{gt}}\ket{0}:=&\sqrt{\frac{\Phi_g(t+\delta t)}{\Phi_g(t)}}\ket{\varsigma_{gt+\delta t}}\ket{0}\nonumber\\
+&\sum_{xg'}\sqrt{\frac{\int_t^{t+\delta t}P(x,g'|g)\phi_{g'g}^x(t')dt'}{\Phi_g(t)}}\ket{\varsigma_{g'0}}{\ket{x}}.
\end{align}

We are now in a position to now generalise Algorithm 1 for such processes. It transpires that this is for the most part simply a case of repeating the steps for renewal processes multiple times for each of the dwell functions.

To generalise Steps 1 and 2, we define a function \mbox{$\psi_{g'g}^x(t):=\sqrt{\phi_{g'g}^x(t)}$} for each of the dwell functions, and analogous to the case of renewal processes, approximate each of them by finite exponential sums $\tilde{\psi}_{g'g}^x(t)$:
\begin{equation}
\tilde{\psi}_{g'g}^x(t)=\sum_{j=1}^Nc_j^{g'gx}e^{(-\gamma_j^{g'gx}+i\omega_j^{g'gx})t}.
\end{equation}

Generalising Steps 3 to 5, we then similarly use these to construct a set of generator states $\{\ket{\varphi_j^{g'gx}}\}$, again defined implicitly in terms of an evolution operator:
\begin{align}
\label{eqgenerallossyu}
\tilde{U}_{\delta t}\ket{\varphi_j^{g'gx}}\ket{0}&=e^{(-\gamma_j^{g'gx}+i\omega_j^{g'gx})\delta t}\ket{\varphi_j^{g'gx}}\ket{0}\nonumber\\
&+\sqrt{1-e^{-2\gamma_j^{g'gx}\delta t}}\ket{\tilde{\varsigma}_{g'0}}\ket{x}.
\end{align}
Here we have analogously defined memory states as linear combinations of these generator states:
\begin{equation}
\label{eqfullapproxmemstates}
\ket{\tilde{\varsigma}_{gt}}\propto\sum_{xg'j}\sqrt{P(x,g'|g)}\frac{c_j^{g'gx}}{\sqrt{2\gamma_j^{g'gx}}}e^{(-\gamma_j^{g'gx}+i\omega_j^{g'gx})t}\ket{\varphi_j^{g'gx}}.
\end{equation}
This implicit definition can be used to determine the overlaps of the generator states, from which a reverse Gram-Schmidt procedure can be used to express them explicitly in terms of (at most) $N|\mathcal{X}||\mathcal{G}|$ orthonormal basis states\footnote{It is possible that some generator states may be linear combinations of those belonging to other modes. This does not break the protocol, though will result in some memory dimensions being left unused. It may then be possible to use these dimensions to incorporate additional terms into the approximate exponential sums to increase their accuracy.}. In turn, the evolution operators and memory states may be expressed in this basis, completing the protocol.

We remark on a useful feature of this compression -- that the modal wait-time distributions maintain their structure as a product of symbolic dynamics and a temporal component -- with only this latter factor modified. That is, the compressed quantum models have the statistics of a process with the same transition topology, but now with modal wait-time distributions $\sum_{xg'}P(x,g'|g)\tilde{\phi}_{g'g}^x(t)$, where $\tilde{\phi}_{g'g}^x(t)=|\tilde{\psi}_{g'g}^x(t)|^2$. This distortion introduces errors only in terms of the times when events occur, and not the probability with which they occur. Moreover, the product structure entails that the distortion in the statistics of the compressed quantum model is no greater than the worst of the distortions of the $\phi_{g'g}^x(t)$, and that the errors in each inter-event interval are independent. Thus, the performance of the protocol seen in the renewal process examples will still hold in this generalised setting. The memory of the resultant quantum model will be compressed to at most $N|\mathcal{X}||\mathcal{G}|$ dimensions.

\subsection{Example}
\label{secgenexample}

\begin{figure}
\includegraphics[width=\linewidth]{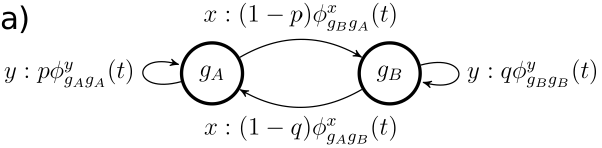}
\includegraphics[width=\linewidth]{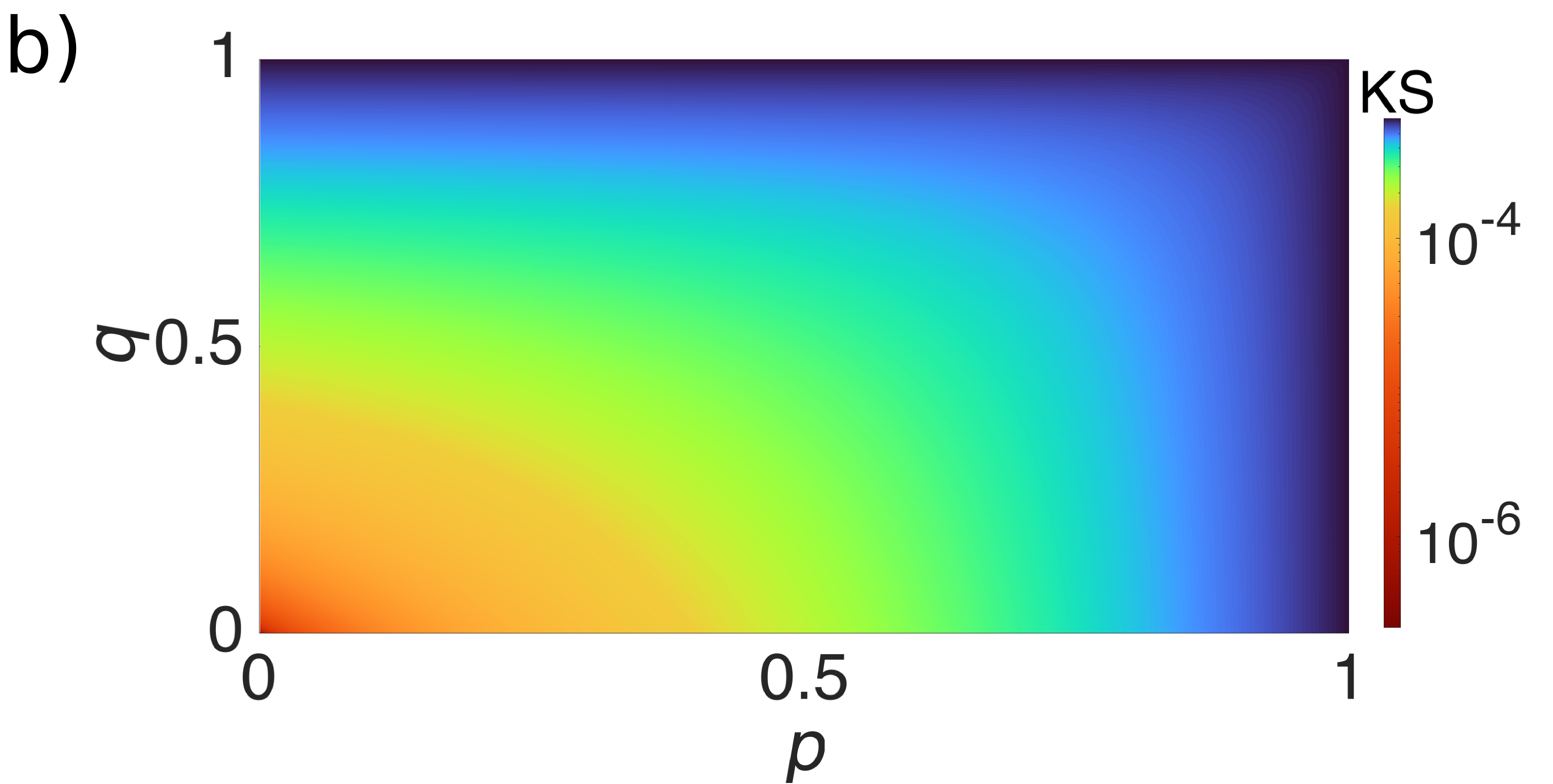}
\caption{(a) HSMM representation of example discussed in Section \ref{secgenexample}. (b) Corresponding averaged KS statistic with a 32-dimensional memory for full $(p,q)$ parameter range.}
\label{figgenexample}
\end{figure}

As an illustration of how the general case is little more than a straightforward application of Algorithm 1 for multiple times, we apply it to an example process consisting of dwell functions that are based on the examples above. Specifically, the process has two modes ${g_A,g_B}$ and two possible events ${x,y}$, with the dwell function of both modes corresponding to an alternating Poisson process for event $x$ and a bimodal Gaussian process for event $y$, and a transition structure such that the mode changes on event $x$ and remains constant on event $y$; the probabilities of each event is different for the two modes. This is depicted as a HSMM in \figref{figgenexample}(a).

We measure the error in the accuracy as the average KS statistic, where the average is taken over events (for simplicity we scale all dwell functions to have the same average mean firing rate, such that this also essentially corresponds to the average over time). That is, the average KS statistic $\bar{\mathrm{KS}}:=\sum_{xgg'}\mathrm{KS}(\phi_{g'g}^x(t),\tilde{\phi}_{g'g}^x(t))P(x,g)$. It is possible to apply the KS statistic in this way as the errors are constrained to a single interevent interval, and there is no crossover of errors between the dwell functions of different events. Moreover, we need not calculate the approximations of the dwell functions anew -- the approximations (and corresponding errors) found in Section \ref{secexamp} are the very same approximations needed. In \figref{figgenexample}(b) we plot this for the full $(p,q)$ parameter range for $N=8$ (requiring 32 memory dimensions in total). Of note are the limits $p=q=0$ (corresponding to only alternating Poisson process) and $p=q=1$ (corresponding to only bimodal Gaussian process) where the errors take on their minimum and maximum respectively, matching with those found for the renewal processes, while the error for the remainder of the parameter space interpolates between these two limits. Note that we neglect the extra dimensions made available by linear dependencies of generator states at the exceptional parameter regimes $p=0,1$, $q=0,1$, and $p=q$.

\subsection{Scope for improvement?}

\begin{figure}
\includegraphics[width=\linewidth]{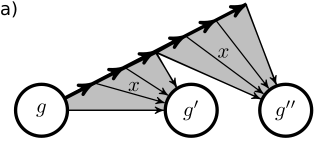}
\includegraphics[width=\linewidth]{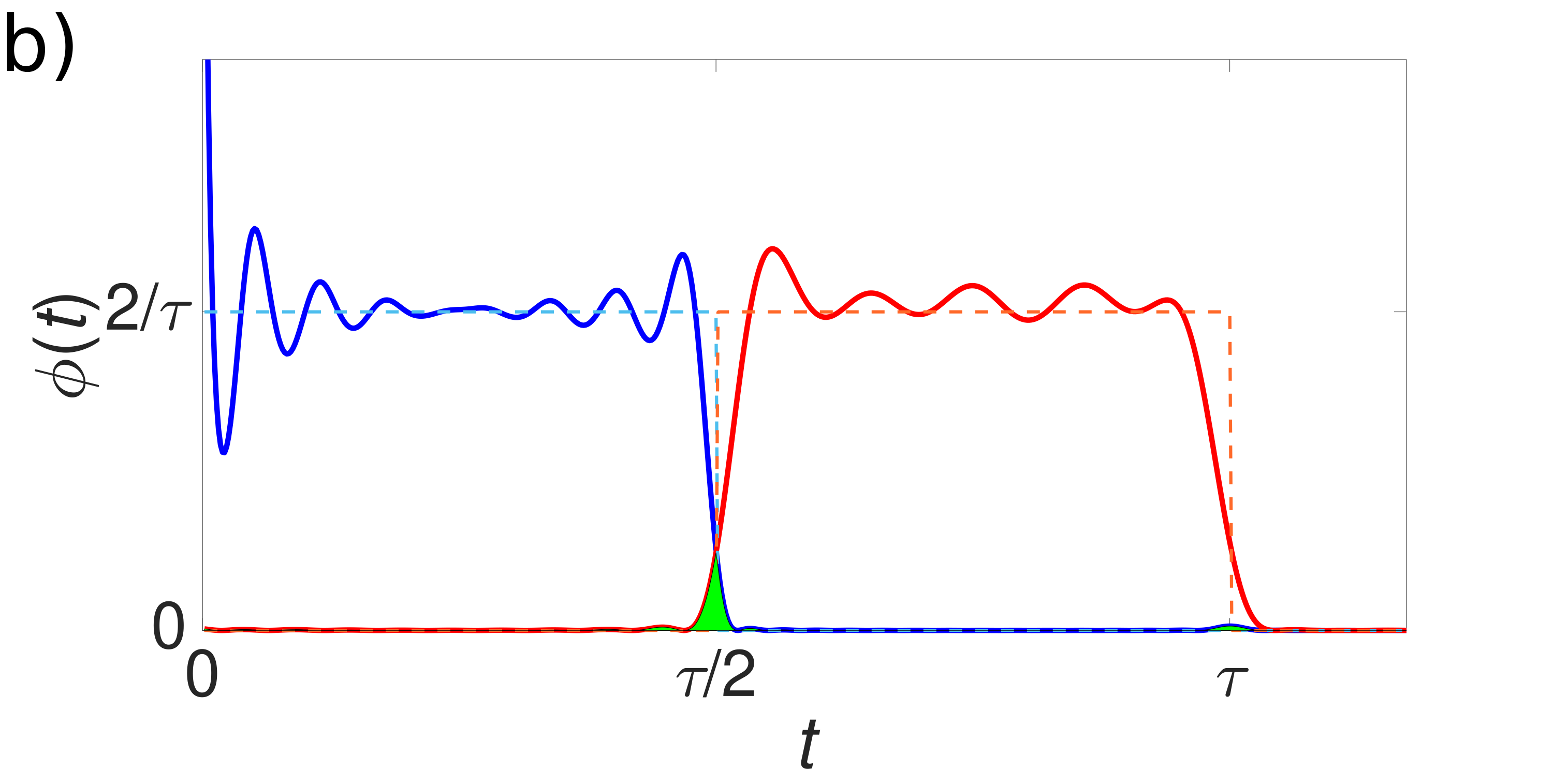}
\caption{(a) HMM representation of a transition that does not satisfy the classical condition on mode update rules. (b) Example of spurious overlap of dwell functions introduced by the compression protocol. Dashed lines show the exact dwell functions, solid lines the approximations, and green the regions of overlap.}
\label{figgenfail}
\end{figure}

Above, we have followed the classical condition that the HSMM representation is such that the symbol and current mode alone determine the next mode. Yet, the quantum models described in Eq.~\eqref{eqexactgeneralu} still function correctly -- and remain causal -- with only the weaker condition on the HSMM representation that the triple $(g,x,t_0)$ suffices to determine the subsequent mode. That is, emission of a given symbol from a given mode can result in a transition to two (or more) possible different modes, provided that also knowing the time spent in the current mode then provides sufficient information to determine the next mode. That is, the classical convention requires $H(G'|G,X)=0$, while the exact quantum models assume only that $H(G'|G,X,T_0)=0$ (here, $H(.)$ is the Shannon entropy~\cite{nielsen2000quantum}). An example of such a transition satisfying only the weaker condition is illustrated in \figref{figgenfail}(a).

However, in the case where only this weaker condition holds, there can be interference between the generator states corresponding to transitions with the same symbol and initial mode, but different end mode. This manifests from errant overlaps of the approximate dwell functions $\tilde{\phi}_{g'g}^x(t)$: while \mbox{$\int_0^\infty\phi_{g'g}^x(t)\phi_{g''g}^{x}(t)dt=0\forall x,g,g',g''\neq g'$}, this may not hold true for the $\tilde{\phi}_{g'g}^x(t)$. That is, there may be times $t$ for which $\tilde{\phi}_{g'g}^x(t)$ and $\tilde{\phi}_{g''g}^x(t)$ \mbox{($g''\neq g')$} are simultaneously non-zero, violating the condition $H(G'|G,X,T_0)=0$. Such a violation cannot occur under the stronger classical condition, as we are already guaranteed there is at most one $g'$ for each pair $(g,x)$ for which $\phi_{g'g}^x(t)$ (and thus $\tilde{\phi}_{g'g'}^x(t)$) is not zero everywhere.

As an example, consider a process where the dwell time associated with mode $g$ and event $x$ is uniformly distributed over the interval $[0,\tau]$, with the system transitioning into mode $g'$ if the dwell time is less than $\tau/2$, and into $g''$ if it is greater than (or equal to) $\tau/2$. Then, $\phi_{g'g}^x(t)$ is a uniform distribution over $[0,\tau/2)$, and $\phi_{g''g}^x(t)$ a uniform distribution over $[\tau/2,\tau]$. When we parse this through the compression protocol, the approximated distributions have a non-zero overlap, and so have interfering probability amplitudes. This is illustrated in \figref{figgenfail}(b) for $N=16$.

This interference requires us to modify the symbolic transition probabilities $P(X,G'|G)$ to an approximate form $\tilde{P}(X,G'|G)$ in order to appropriately normalise the memory states, which will correspondingly distort the transition structure. In particular, it can result in the model transitioning to superpositions of memory states, manifesting new (potentially infinitely many) effective modes. While these effective modes do not require additional memory dimensions to track (as they are linear combinations of existing memory states), they do allow for a gradual accumulation of errors over time, as the errors are now able to propagate across multiple inter-event intervals. A further complication is presented in the freedom of choice in how to actually assign $\tilde{P}(X,G'|G)$ to enforce proper normalisation -- while a simple rescaling of $P(X,G'|G)$ would work, it is also possible to achieve this with an uneven rescaling, which may result in greater accuracy by offsetting the effect of the interference.
 
Note that the magnitude of the interference scales with the overlaps of the memory states for each mode -- and hence the overlaps of their statistics: thus, the more distinguishable the statistics of the modes, the smaller the distortion. Further, as noted above, with the stronger condition imposed on classical models these overlaps cannot occur, and thus when compressing a given such classical model we can sidestep such interference. Nevertheless, embracing this weaker condition may unlock even greater compression potential; we leave the optimisation of the $\tilde{P}(X,G'|G)$ in such settings as an open question for future work. 

\section{Discussion}

We have introduced a lossy compression protocol for the quantum modelling of stochastic temporal dynamics. By harnessing non-classical features of quantum state spaces -- namely, that sets of quantum states can be at once linearly-dependent and non-degenerate -- an effective coarse-graining of the state space inhabited by a quantum memory can be realised. This achieves a much greater compression than is possible with analogous classical methods and exact quantum compression alike. The relaxation from exact to near-exact replication naturally fits into applications where the dynamics of the system to be modelled have been inferred through observation~\cite{marzen2020inference, ho2020robust}, and are thus already an approximation of the true dynamics. This also brings the additional benefit of placing less  demand on the precision of the quantum processor implementing the simulation, which in current realistic settings should not be assumed noiseless. 

Going forwards, our work encourages the development of similar lossy compression beyond tracking the temporal component of stochastic processes. For example, the framework can be applied to compress quantum clocks~\cite{woods2018quantum, yang2020ultimate}, and motivates the extension to other models with continuous state spaces, such as belief spaces~\cite{cassandra1994acting, kaelbling1996reinforcement} used in reinforcement learning~\cite{sutton2018reinforcement}. Further avenues include development of analogous methods for compressed modelling of purely symbolic dynamics and input-output processes~\cite{thompson2017using}. 

Furthermore, in spite of the significant compression advantage offered by our protocol, it is by no means optimal. Two aspects we foresee as presenting opportunities for enhancing the compression are in the choice of algorithm for constructing an approximate exponential sum, and in allowing for more general complex $\psi(t)$ to be considered. Pursuing the former of these may allow for more faithful approximations of the wait-time distribution without increasing the number of allowed states. In the latter we have a family of functions we can attempt to approximate, and we need only take the one which we can most faithfully represent. In the case of general continuous-time processes, the question remains open how to best handle cases where the classical condition that the dynamics factor into a product of temporal and symbolic dynamics does not hold. Further improvements in this regime may also be found by taking a more holistic approach that coarse-grains the Hilbert space in terms of symbolic and temporal dynamics simultaneously. Nevertheless, even in this initial foray, we see the potential for drastic improvement over classical techniques. Moreover, the high fidelities reached with comparatively few dimensions places it well within reach of current and near-term small-scale quantum processors with only a handful of qubits~\cite{negnevitsky2018repeated, ghafari2019dimensional}, offering exciting prospects for imminent experimental realisations.

\appendix

\section{Approximate exponential sums}
\label{secdecomp}

In Algorithm 1, Step 2 requires that we construct an exponential sum approximating the square root of the wait-time distribution. Here, we use the method of Beylkin and Monz\'{o}n~\cite{beylkin2005approximation}, summarised in Algorithm 2.

\begin{algorithm}[H]
\caption{\textsf{\\\mbox{Approximate exponential sum}~\cite{beylkin2005approximation}}}
\begin{flushleft}
\emph{Inputs}: Exact function $\psi(t)$, with the domain scaled such that the region to be approximated is the interval $[0,1]$, target precision $\epsilon$. \\
\emph{Outputs}: Set of triples $\{(c_j,\gamma_j,\omega_j)\}$ yielding approximate decomposition $\psi(t)\approx\sum_{j=1}^Nc_j\exp((-\gamma_j+i\omega_j)t)$.
\end{flushleft}
\begin{algorithmic}[1]
\STATE Construct $M$-dimensional vector $h_j:=\psi(j/M)$ for $0\leq j\leq M$ with $M$ sufficiently large to oversample function.
\STATE Construct Hankel matrix $H_{jk}:=h_{j+k}$ and find eigenvector $\sigma$ corresponding to the eigenvalue closest to $\epsilon$.
\STATE From the elements of $\sigma$, construct the polynomial $\sum_{j=0}^{M/2}\sigma_jz^j$ and solve to find the roots $\{\Gamma_k\}$.
\STATE Construct Vandermonde matrix $V_{jk}:=\Gamma_k^j$ for \mbox{$0\leq j \leq M$}. Invert to find solutions \mbox{$\{c_k\}$ to $h_j:=\sum_{k=1}^{M/2}c_kV_{jk}$}.
\STATE Define $\{-\gamma_j\}$ and $\{\omega_j\}$ as the real and imaginary parts of $\{\ln(\Gamma_j)\}$ respectively.
\end{algorithmic}
\end{algorithm}

For our purposes, to obtain an $N$ term approximate sum we keep only the triples with the $N$ largest magnitudes for weights $\{|c_j|\}$. Prior to this we also discard any terms with non-positive $\gamma_j$ (to ensure a valid quantum state of the form Eq.~\eqref{eqquantumrenewalstates} can be constructed), and rescale the weights by a constant factor to ensure the sum has unit $L_2$ norm. By varying the precision $\epsilon$ we obtain different decompositions, with truncation to fewer terms favouring larger $\epsilon$, and conversely, larger number of terms performing better with smaller $\epsilon$. In our examples, we took $M=1000$ and varied $\epsilon$ to find the most accurate decomposition for each $N$ (according to the KS statistic), ultimately using values in the range $10^{-12}$ to $10^{-1}$. We used GNU Octave's \texttt{roots} function~\cite{octaveroots} to numerically solve the polynomials, and $c=(V^TV)^{-1}V^Th$ to solve the overconstrained Vandermonde system. 

\section{Lossy classical compression method}
\label{secclassicalcomp}

\begin{figure}
\includegraphics[width=\linewidth]{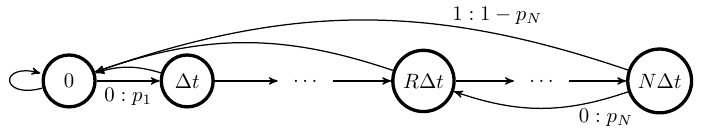}
\caption{HMM topology of the most accurate $N+1$-dimensional approximation of a renewal process. The variable parameters are the the transition probabilities $\{p_j\}$, the timestep size $\Delta t$, and the position of the loop $R$.}
\label{figapproxrenewalhmm}
\end{figure}

As shown in \figref{figrenewalhmm}, the transition structure between the memory states of the $\varepsilon$-machine of a renewal process takes the form of an incrementing counter that resets upon events. A finite-dimensional approximation must adopt the structure of \figref{figapproxrenewalhmm}, where the counter progresses up to a terminal state, upon which it loops back to an earlier state~\cite{elliott2020extreme}. The variable parameters are the transition probabilities $\{p_j\}$, the timestep size $\Delta t$, and the target state of the loop $R$.

The optimal lossy classical compression at fixed dimension is found by minimising the associated cost function over all possible choices of these parameters. We use a standard gradient descent-based approach to seek the minimum of the KS statistic. For each possible choice of loop state $R$, we generate $W$ seeds of random parameters for $(\{p_j\},\Delta t)$ and run $S$ steps of update according to \mbox{$p_j\to p_j-\eta_p\del_{j} D(\{p_j\},\Delta t)$} and \mbox{$\Delta t\to\Delta t-\eta_t\del_t D(\{p_j\},\Delta t)$}, where $D$ is the KS statistic (with hard constraints to ensure the parameters remain physical). We then keep the final parameter set that reached the minimum value of $D$ across all choices of loop state and seeds. As with the quantum method we rescaled the wait-time distribution to the domain $[0,1]$, and for purposes of numerical evaluation discretised it into 1000 steps. We again remark that we do not claim this method to necessarily yield the very optimal lossy classical compression at fixed dimension, but simply that it should offer a ballpark figure as to its performance. That is, we believe it is reasonable to expect the optimal classical compression will not perform significantly better than the explicit examples we find here.

We generated the initial seeds for $\{p_j\}$ uniformly in the interval $[0,1]$, and $\Delta t$ exponentially decaying. For the alternating Poisson process we found best performance by taking learning rates $\eta_p=10^{-4}$ and $\eta_t=10^{-8}$, with gradients estimated over discrete intervals $\delta p=10^{-3}$ and $\delta t=10^{-4}$. Empirically, the descents appeared to converge on a minimum within $S=12500N$ steps, and running more than $W=1000$ seeds for each loop state did not seem to yield any improved minima. For the bimodal Gaussian process we found best perfomance with much the same parameters; a slight improvement was found by increasing the learning rates by a factor of 10 for the first $1250N$ steps of descent, whereupon convergence was reached within $S=6250N$ steps.

\section{Compression of top-hat distributions}
\label{sectophat}

In Section \ref{seccostly} we discussed how wait-time distributions with sharp peaks are hard to compress. Here we illustrate this with a case study of renewal processes with top-hat distributions, showing how performance of the quantum compression degrades with narrowing of the width.

Such top-hat distributions of width $\Delta t$ take on uniform value between $\tau-\Delta t$ and $\tau$ (and zero elsewhere), with $\tau$ forming a arbitrary scaling factor. We consider the cases $\Delta t/\tau=2^0,2^1,\ldots,2^5$, and used Algorithm 1 to construct quantum models of between one and five qubit memories. In running Algorithm 2 we set $M=6000$ and placed $\tau$ at 512, with the long timescale properly accounting for the long tails of the poorer-performing models. Best performance was found for $\epsilon$ in the range of $10^{-3}$ to $10^2$.

\begin{figure}
\includegraphics[width=0.91\linewidth]{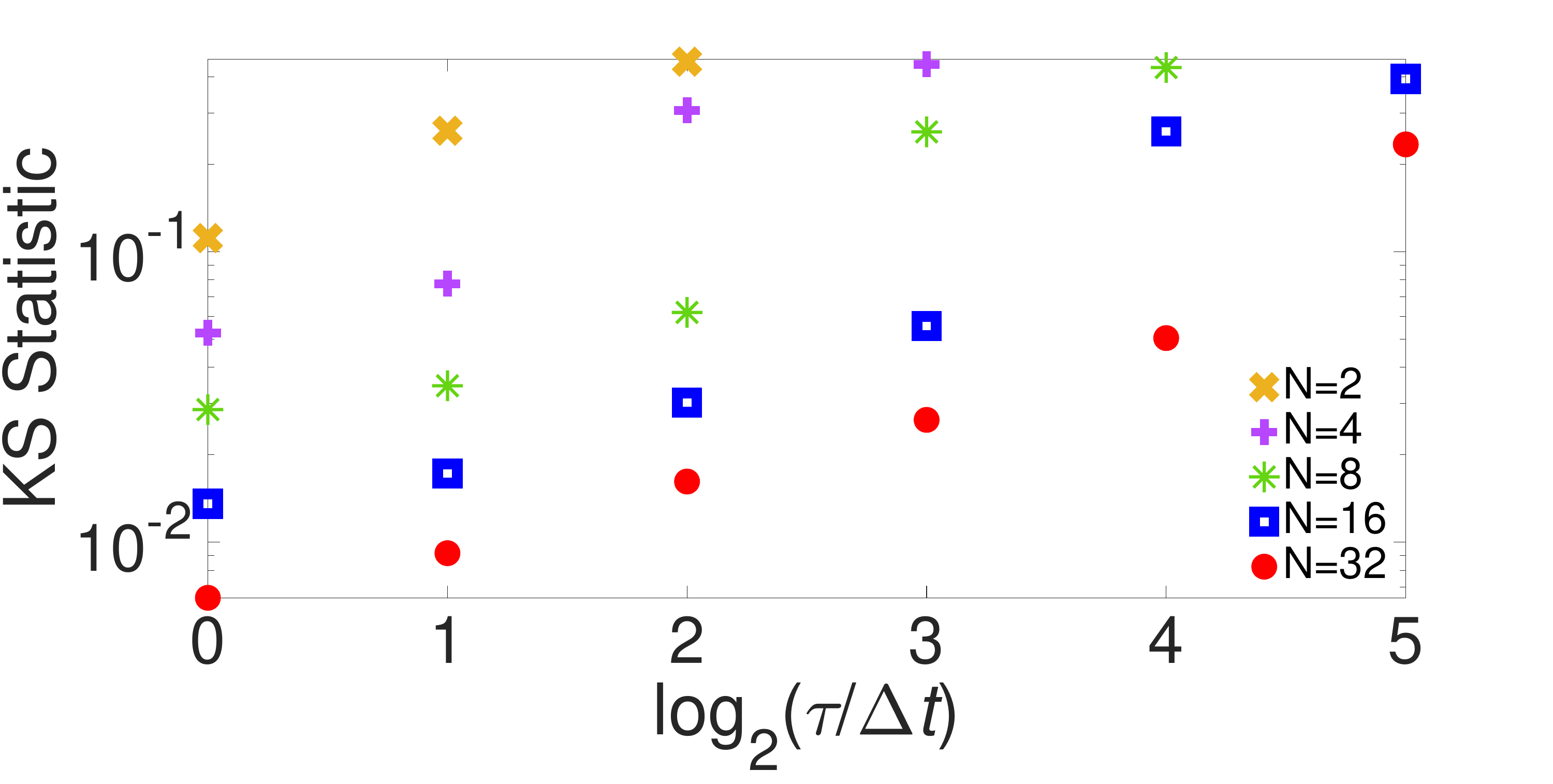}
\caption{KS statistics showing performance of compressed quantum models for renewal processes with top-hat distributions of varying width.}
\label{figtophatks}
\end{figure}

\begin{figure}
\includegraphics[width=\linewidth]{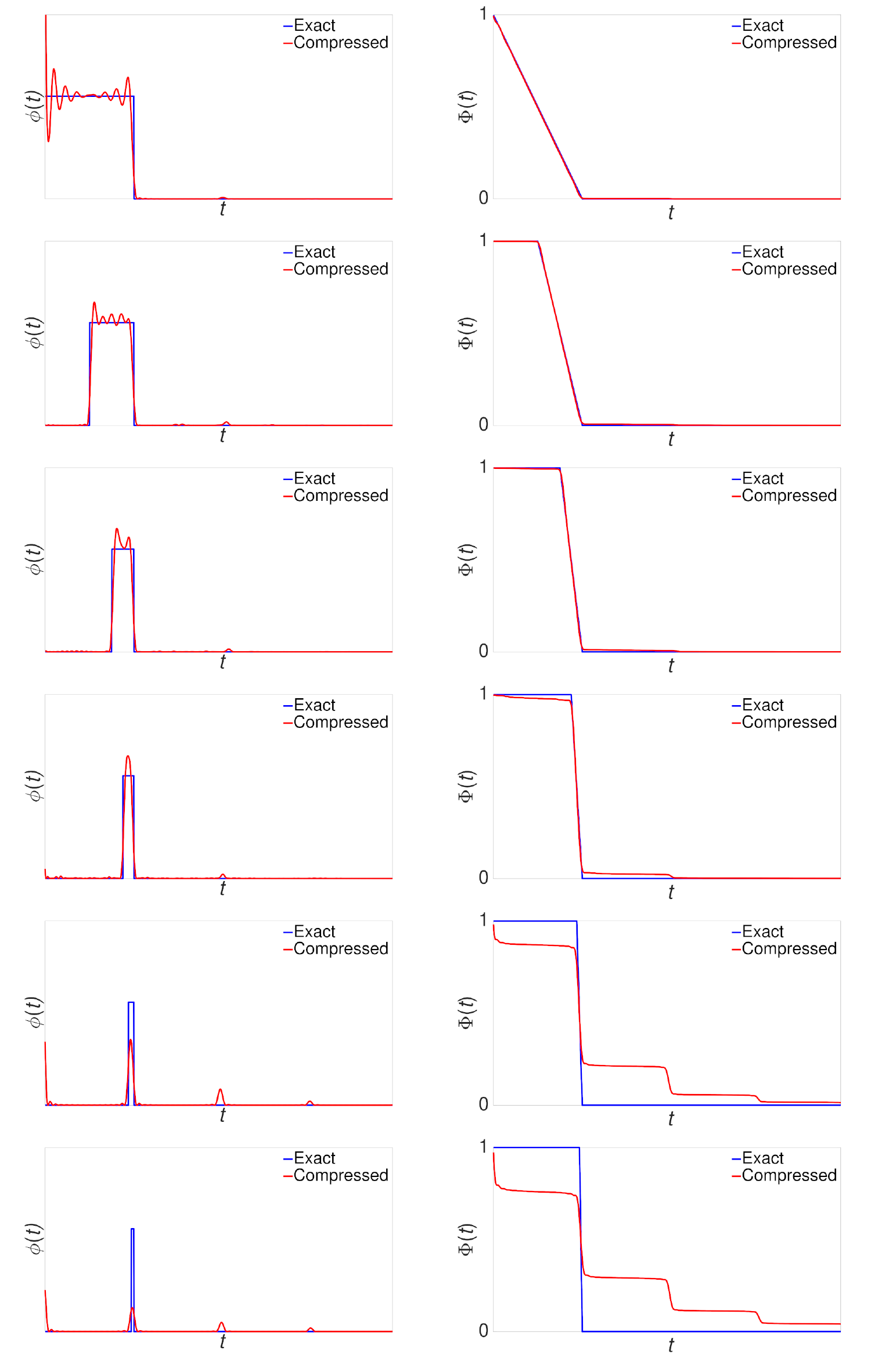}
\caption{Wait-time distributions (left) and survival probabilities (right) of four-qubit compressed quantum models models for renewal processes with top-hat distributions of varying width (arbitrary units).}
\label{figtophatplots}
\end{figure}

In \figref{figtophatks} we plot the KS statistic found for the quantum models, truncated to the smallest width the model could simulate with a KS statistic below $~0.45$ (note that a memoryless model of any process exists with a KS statistic no greater than 0.5). We see that each halving of the width requires an additional qubit to model with roughly the same accuracy, as one would intuitively expect -- in the classical case, doubling the number of states allows a model with half the timestep size. As a point of comparison, consider that a classical model with $N<\tau/\Delta t$ cannot beat a KS statistic of 0.5; to see this, consider that the best classical model in this instance would be a deterministic counter that emits only on the last state, positioned to coincide with the time where $\Phi(t)=0.5$. In \figref{figtophatplots} we compare wait-time distributions and survival probabilities of the compressed four-qubit quantum models to their exact counterparts for each of the widths. As the width narrows, the periodicity of the approximate distributions can be seen, due to competition between surpressing these spurious peaks with the exponential decay, and the need to not supress the modelled peak.

\acknowledgments
We thank Andrew Garner and Mile Gu for discussions. This work was funded by the Imperial College Borland Fellowship in Mathematics and grant FQXi-RFP-1809 from the Foundational Questions Institute and Fetzer Franklin Fund (a donor advised fund of the Silicon Valley Community Foundation).

\pagebreak
\bibliography{ref}

\end{document}